\documentclass[aps,jcp,preprint,a4paper,onecolumn,unsortedaddress]{revtex4}

\usepackage[fleqn]{amsmath}
\usepackage{amssymb}
\usepackage[dvips]{graphicx}
\usepackage{color}
\usepackage{tabularx}
\usepackage{algpseudocode}
\usepackage{multirow}


\renewcommand{\vec}[1]{{#1}}
\newcommand{\recheck}[1]{{\color{black} #1}}

\newcommand{\nb}{\textrm{band}}
\newcommand{\pos}[1]{\vec{R}\,(#1)}
\newcommand{\posi}[2]{\vec{R}_{#1}(#2)}
\newcommand{\dpos}[1]{\Delta\vec{R}_{#1}}
\newcommand{\ket}[1]{|#1\rangle}
\newcommand{\ala}[0]{{\textrm{A}}}
\newcommand{\alm}[0]{{\textrm{M}}}
\newcommand{\scf}[0]{{\textrm{scf}}}
\newcommand{\xpl}[0]{{\textrm{xpl}}}
\newcommand{\tx}[0]{{\textrm{T}}}
\newcommand{\gx}[0]{{\textrm{G}}}

\begin{document}

\title{On the existence of the optimal order for wavefunction extrapolation in Born-Oppenheimer molecular dynamics}
\author{Jun Fang}
\affiliation{Institute of Applied Physics and Computational Mathematics}
\affiliation{CAEP Software Center for High Performance Numerical Simulation}
\author{Xingyu Gao}
\affiliation{Laboratory of Computational Physics}
\affiliation{Institute of Applied Physics and Computational Mathematics}
\affiliation{CAEP Software Center for High Performance Numerical Simulation}
\author{Haifeng Song}
\affiliation{Laboratory of Computational Physics}
\affiliation{Institute of Applied Physics and Computational Mathematics}
\affiliation{CAEP Software Center for High Performance Numerical Simulation}
\author{Han Wang}
\email{wang_han@iapcm.ac.cn}
\affiliation{Institute of Applied Physics and Computational Mathematics}
\affiliation{CAEP Software Center for High Performance Numerical Simulation}
   
\begin{abstract}
  Wavefunction extrapolation greatly reduces the number of self-consistent field (SCF) iterations and thus the overall computational cost of
  Born-Oppenheimer molecular dynamics (BOMD) that is based on the Kohn--Sham density functional theory. 
  Going against the intuition that the higher order of extrapolation possesses a better accuracy, 
  we demonstrate, from both theoretical and numerical perspectives,
  that the extrapolation accuracy firstly increases and then decreases with respect to the order,
  and an optimal extrapolation order in terms of minimal number of SCF iterations always exists.
  We also prove that the optimal order tends to be larger
  when using larger MD time steps or more strict SCF convergence criteria.
  By example BOMD simulations of a solid copper system, we show that the optimal extrapolation order
  covers a broad range when varying the MD time step or the SCF convergence criterion.
  Therefore, we suggest the necessity for BOMD simulation packages to
  open the user interface and to provide more choices on the extrapolation order.
  Another factor that may influence the extrapolation accuracy is the alignment scheme
  that eliminates the discontinuity in the wavefunctions with respect to the atomic or cell variables.
  We prove the equivalence between the two existing schemes,
  thus the implementation of either of them does not lead to essential difference in the extrapolation accuracy.  
\end{abstract}

\maketitle

\section{Introduction}
Among \textit{ab-initio} molecular dynamics (AIMD) simulation methods,
Born--Oppenheimer molecular dynamics (BOMD) is considered to
have good reliability and predictive power, especially for
material systems with a small or vanishing band gap
\cite{marx-hutter00,bornemann98,tangney06},
and thus has been accepted as the default choice in several well-known AIMD
simulation packages \cite{kresse93,kresse96cms,qe09,abinit09,castep05}.
BOMD involves solving the quantum equation
(for us, the Kohn--Sham equation \cite{hk64,ks65,martin04})
for each instantaneous atomic configuration to obtain the atomic and cell forces,
and propagating the atomic and cell variables
according to classical Newtonian dynamics \cite{marx-hutter00}.
The computational cost of this very time-consuming procedure
can be strongly reduced by the extrapolation technique
\cite{marx-hutter00,niklasson06},
which combines results from previous MD time steps to predict
a good initial guess for the self-consistent field (SCF) solution of
the Kohn--Sham equation at the current MD time step.
In the plane-wave method, the Kohn--Sham Hamiltonian matrix is never assembled explicitly
and generally diagonalized by iterative schemes~\cite{payne92,kresse96cms,martin04},
so both the extrapolation of electron density and
that of the Kohn--Sham eigenvalues and wavefunctions/orbitals are necessary.
In this work, we discuss only the wavefunction extrapolation because
it covers all difficulties in the extrapolation technique.

The existing extrapolation schemes could be classified into two categories
\cite{zheng11}.
The first is the time extrapolation (TX),
which directly extrapolates the wavefunctions as functions of time.
It can be implemented either by
the Lagrange's polynomial extrapolation formula~\cite{arias92prl,vandevondele05,atsumi08},
or by the polynomial least-squares fitting 
proposed by literature on the Fock matrix dynamics~\cite{pf04,herbert05}.
In the following text, we refer to the first implementation when we use the terminology ``TX''.
The second category is the geometric extrapolation (GX),
which determines the wavefunction extrapolation coefficients by minimizing the distance
between the current atomic coordinates and a linear combination of 
atomic coordinates from the previous time steps~\cite{arias92prb,alfe99,atsumi10}.
These schemes have been implemented 
in several popular AIMD simulation packages using the Kohn--Sham density functional theory
and a plane-wave basis set, see Tab.~\ref{tbl:pkginfo} for an incomplete summary.
It is worth noting that the order of the TX scheme usually refers to the order of the polynomial used to fit the wavefunctions,
but the meaning of ``order'' in the context of GX scheme is not very clear.
However, both TX and GX schemes can be uniformly characterized by the number of wavefunctions
from previous MD time steps used in the extrapolation.
Therefore, we accept this convention and denote the number by $M$ in the rest of the manuscript.

\begin{table}[]
    \centering
    \caption{Wavefunction extrapolation provided in several popular AIMD simulation packages
    using the Kohn--Sham density functional theory and a plane-wave basis set.
    TX and GX refer to time extrapolation with Lagrange polynomial
    and geometric extrapolation, respectively.
    $M$ is the number of used MD time steps in the extrapolation.
    APJ and Mead denotes the wavefunction alignment scheme proposed by Arias, Payne and Joannopoulos,
    and the scheme from Mead, respectively.}
    \label{tbl:pkginfo}
    {\setlength{\tabcolsep}{1em}
    \begin{tabular}{l|c|c|c}
        \hline
        Package & Extrapolation scheme & $M$ & Alignment scheme \\\hline
        VASP \cite{kresse93,kresse96cms} & GX     & 3    & APJ  \\
        ABINIT \cite{abinit09}           & GX     & 3    & APJ  \\
        Quantum Espresso \cite{qe09}     & GX     & 3    & Mead \\
        CASTEP \cite{castep05}           & GX, TX & 2, 3 & APJ  \\
        \hline
    \end{tabular}}
\end{table}

The effectiveness of the extrapolation is calibrated by the number of iterations used to
achieve a converged SCF solution under some user-defined convergence criterion.
At first glance, the scheme using more previous-step information is more effective in predicting the wavefunctions.
However, in practice, the number of SCF iterations was observed to firstly decrease
and then to increase when extrapolating from an increasing number of previous wavefunctions~\cite{pf04,herbert05}.
That is to say, an optimal value of $M$ that minimizes the number of SCF iterations was observed in these numerical tests.
Moreover, it was reported,
the optimal value of $M$ gets larger when tightening the SCF convergence criterion~\cite{herbert05}.
These phenomena, as far as we see, lack proper theoretical explanation~\cite{steele10}.
Most of the well-established AIMD packages provide very limited choices on the value of $M$ (see Tab.~\ref{tbl:pkginfo}),
and it is still questionable whether these configurations are optimal in terms of
the number of SCF iterations.

In this work, we present an error analysis for the TX scheme
in terms of the MD time step size and the SCF convergence criterion.
Based on the error analysis, we claim that the optimal value of $M$ always exists,
and explain why the optimal $M$ increases when using larger time step sizes or more strict SCF convergence criteria.
Moreover, we theoretically elucidate the similarity between TX and GX schemes,
so the theoretical findings on the TX scheme can be safely extended to the GX scheme.
As numerical examples, we select a solid copper system with 64 atoms,
and carry out microcanonical simulations under different time step sizes and SCF convergence criteria.
The appearance of optimal values of $M$ well validate the error analysis,
and a comparison between TX and GX confirms their similarity.
We demonstrate that the optimal value of $M$ takes 2 in some cases, and takes as large as 5 in other cases.
Therefore, we may suggest the necessity for AIMD simulation packages to
open the user interface and provide more choices on the parameter $M$.

It should be noted that the extrapolations are linear combinations of wavefunctions from several adjacent time steps.
An important underlying assumption, which is usually not satisfied in practice,
is the continuity of these wavefunctions with respect to the atomic or cell variables~\cite{mead92,arias92prb,pancharatnam56,berry87}.
Therefore, a prerequisite step for the extrapolation is removing this discontinuity
by using the wavefunction alignment scheme
proposed either by Arias, Payne and Joannopoulos (APJ)~\cite{arias92prb}, or by Mead~\cite{mead92}.
Table~\ref{tbl:pkginfo} summarizes their implementations in the AIMD packages.
We show that the corresponding extrapolated wavefunctions by using these two alignment schemes
differ only up to a unitary transformation,
so a particular choice between them does not lead to substantial difference in the extrapolation accuracy. 

The remained part of this paper is organized as follows:
The TX and GX schemes with an arbitrary value of $M$ are given in Sec.~\ref{sec:txgx}.
Then Sec. \ref{sec:err} presents an error analysis for the TX scheme.
As a prerequisite for the extrapolation, the wavefunction alignment is discussed in Sec.~\ref{sec:align}.
In Sec. \ref{sec:test}, we validate our theoretical analysis by
a set of microcanonical simulation tests on a solid copper system.
Finally, the conclusion is made in Sec. \ref{sec:sum}.

\section{Wavefunction extrapolation schemes}\label{sec:txgx}
We consider the following extrapolation scheme,
which is a linear combination of wavefunctions from $M$ previous MD time steps, i.e.~$t_{n-M+1}, \cdots, t_{n-1}, t_n$, 
and provides an initial guess for the SCF iteration at the ($n+1$)-th time step:
\begin{equation}
    \label{eq:xtrpl-form}
    X^{\mathrm{p}}(\pos{t_{n+1}}) =
    X(\pos{t_n}) + \sum_{k=1}^{M-1} c_k \left[
    X(\pos{t_{n-k+1}}) - X(\pos{t_{n-k}}) \right],
\end{equation}
where $X$ represents any of the $N_\nb$ wavefunctions,
the superscript ``p'' indicates the predicted function,
$\{c_k\}$ are extrapolation coefficients,
and atomic positions $\pos{t}\equiv(\posi{1}{t}, \ldots, \posi{\mathrm{natom}}{t})$
with row vector $\posi{I}{t}$ denoting coordinate of
the $I$-th atom at time $t$.

\subsection{The time extrapolation scheme}
\label{sec:tpoly}
The TX scheme fits the wavefunctions from $M$ previous time steps, 
by an $(M-1)$-th order Lagrange polynomial~\cite{nr07},
and evaluates the polynomial at the $(n+1)$-th time step to provide the initial guess for the SCF iteration.
If a fixed time step is used,
then the extrapolation coefficients $\{c_k\}$
are computed by
\begin{equation}
  \label{eq:tpoly}
  c^\tx_k = (-1)^{k-1} \binom{M-1}{k},\quad k = 1,2,\ldots,M-1,
\end{equation}
where $\binom{M-1}{k}\equiv\frac{(M-1)!}{(M-1-k)!\,k!}$.
The superscript ``$\tx$'' means that the coefficients are derived from TX scheme.
Table \ref{tbl:tpoly} lists extrapolation coefficients of schemes with $M$ from 2 to 5.
It is noted that the extrapolation coefficients can be easily computed for simulations with varying time steps.

\begin{table}[]
    \centering
    \caption{Extrapolation coefficients of the TX scheme with $M$ from 2 to 5.}
    \label{tbl:tpoly}
    {\setlength{\tabcolsep}{1em}
    \begin{tabular}{ccccc}
        \hline
        $M$ & $c^\tx_1$ & $c^\tx_2$ & $c^\tx_3$ & $c^\tx_4$ \\
        \hline
        2 & 1       & --      & --    & --      \\
        3 & 2       & $-1$    & --    & --      \\
        4 & 3       & $-3$    & 1     & --      \\
        5 & 4       & $-6$    & 4     & $-1$    \\
        \hline
    \end{tabular}}
\end{table}

\subsection{The geometric extrapolation scheme}\label{sec:rmin}
The extrapolation coefficients
$\{c_k\}$ of the GX scheme are determined by minimizing the distance
between atomic coordinates of the ($n+1$)-th time step
and the extrapolated atomic coordinates from $M$ previous time steps:
\begin{equation}
    \label{eq:rmin-obj}
    \min_{\{c_k\}}
    g(\{c_k\})
    \equiv
    \min_{\{c_k\}}\left\vert \posi{}{t_{n+1}} - \vec{R}_{}' \right\vert^2,
\end{equation}
where the extrapolated atomic coordinates $\vec{R}'$ are given by
\[
\vec{R}_{}'
\equiv
\posi{}{t_n} + \sum_{k=1}^{M-1} c_k \left[\posi{}{t_{n-k+1}} - \posi{}{t_{n-k}} \right].
\]
The solution of the minimization problem (\ref{eq:rmin-obj})
with respect to the extrapolation coefficients $\{c_k\}$
is given by the solution of the following linear system:
\begin{align}\label{eq:rmin-ls}
  A c = b,   
\end{align}
where $A$ is a $(M-1)\times(M-1)$ matrix, the elements of which are
\begin{align}
  A_{k_1,k_2} = \dpos{n-k_1+1}\cdot\dpos{n-k_2+1}, \quad 1\leq k_1,k_2\leq M-1,
\end{align}
and $b = (b_1, b_2, \cdots, b_{M-1})^\top$ is a column vector defined by
\begin{align}
  b_k = \dpos{{n-k+1}}\cdot\dpos{{n+1}}, \quad 1\leq k\leq M-1,
\end{align}
with
$
\dpos{j} \equiv \posi{}{t_j}- \posi{}{t_{j-1}}.
$
The solution of the optimization problem~\eqref{eq:rmin-obj} is denoted by $c_k^\gx$.

\vskip .3cm
\noindent
\textbf{Remark}: When $R(t)$ is exactly an ($M-1$)-th order polynomial function with respect to time variable $t$,
then the extrapolation coefficients of the GX scheme are identical to
those of the TX scheme, i.e.~$c_k^\gx = c_k^\tx, \ k=1, \cdots, M-1$, because it is obvious that $g(c_k^\tx) = 0$.
In practice, the $R(t)$ is not likely to be a polynomial, however,
it can be very close to a polynomial in the time interval $[t_{n-M+1}, t_{n+1}]$ for sufficiently small time steps.
For practically used time steps in AIMD simulations,
we observe that the extrapolation coefficients of the TX and GX schemes are very similar.
Therefore, the theoretical results derived for the TX scheme can be safely used to analyze the numerical phenomena of the GX scheme.
When the $R(t)$ is close to a polynomial with an order strictly lower than $M-1$,
the linear system~\eqref{eq:rmin-ls} will be underdetermined, and has infinitely many solutions.
What is observed in simulations is that the determinant of matrix $A$ becomes even smaller than the machine precision.
In this case we do not solve the system~\eqref{eq:rmin-ls}, but simply let $c_k^\gx$ to be the same as the TX coefficients.

\section{Error analysis of the wavefunction extrapolation}\label{sec:err}
In practice the SCF procedure is always stopped at a finite number of iterations
by some precision criterion (denoted by $\epsilon_\scf$),
so the not-fully-converged SCF procedure inevitably introduces error in the wavefunctions. 
We denote this error by $E(\pos{t})$, and the wavefunction used in the simulations is the composition of two parts:
\begin{align}
  X(\pos{t}) = \tilde X(\pos{t}) +  E(\pos{t}),
\end{align}
where $\tilde X(\pos{t})$ denotes the fully converged wavefunctions derived from the SCF procedure stopped at an arbitrarily high precision.
Now we compute the difference between the predicted and fully converged wavefunctions at step $t_{n+1}$ by
\begin{align} \label{eq:error-decompose}
  X^{\mathrm{p}}(\pos{t_{n+1}}) - \tilde X (\pos{t_{n+1}})
  =
  \mathcal E^\xpl +   \mathcal E^\scf,
\end{align}
in which the error of the wavefunction prediction is also composed by two parts, which are
\begin{align}\label{eq:decomp-error-xpl}
  \mathcal E^\xpl
  = & \,
  \Big\{ \tilde X(\pos{t_{n}}) + \sum_{k=1}^{M-1} c_k \left[\tilde X(\pos{t_{n-k+1}}) - \tilde X(\pos{t_{n-k}}) \right] - \tilde X(\pos{t_{n+1}})  \Big\}, \\\label{eq:decomp-error-scf}
  \mathcal E^\scf
  = &\,
  \Big\{ E(\pos{t_{n}}) + \sum_{k=1}^{M-1} c_k \Big[E(\pos{t_{n-k+1}}) - E(\pos{t_{n-k}}) \Big] \Big\}.
\end{align}
The first term on the R.H.S.~of Eq.~\eqref{eq:error-decompose} is the error introduced by approximating the fully converged wavefunction at step $t_{n+1}$ by
extrapolating the fully converged wavefunctions at steps $\{t_{n}, \cdots, t_{n-M+1}\}$.
We assume that the fully converged wavefunctions are smooth with respect to $t$,
in the sense that they are infinitely differentiable in the time domain.
For the Lagrange extrapolation, this error is expressed by
\begin{align}\label{eq:error-xpl}
  \mathcal E^\xpl
  =
  \frac{1}{M!}
  \frac{d^M\tilde X(\pos{t})}{d\,t^M}\Big\vert_{t=\xi}
  \prod_{k=0}^{M-1} (t_{n+1} - t_{n-k})
\end{align}
for some $\xi \in [t_{n-M+1}, t_{n+1}]$. In the equation, ${d^M\tilde X(\pos{t})}/{dt^M}\vert_{t=\xi}$
is the $M$-th order derivative of the wavefunction with respect to $t$ and evaluated at $t=\xi$.
Now if it is further assumed that the derivatives are uniformly bounded by some constant $\vert {d^M\tilde X(\pos{t})}/{d\,t^M} \vert \leq C$, then
the extrapolation error is upper bounded by
\begin{align}
  \vert \mathcal E^\xpl \vert \leq C \Delta t^M.
\end{align}
The second term  $\mathcal E^\scf$ on the R.H.S.~of Eq.~\eqref{eq:error-decompose} is the extrapolation of the SCF error.
It should be noted that the SCF error is, in general, not a continuous function with respect to time.
Therefore an estimate of $\mathcal E^\scf$ like Eq.~\eqref{eq:error-xpl} would NOT hold  for any $M\geq 1$.
This means that the Lagrange extrapolation of the SCF error would not be able to predict its value at future times.
Actually, if the SCF error at different time steps are assumed to be independent random variables with identical distribution,
we claim that (details of the proof are provided in Appendix~\ref{app:prove-error-scf})
the root mean-square magnitude of $ \mathcal E^\scf$ grows exponentially with respect to
the value of $M$, and has an estimate of 
\begin{align}\label{eq:error-scf}
  \sqrt{\langle \vert\mathcal E^\scf\vert^2\rangle} > \frac{2^{M-1}}{\sqrt {M}} \sigma_\scf,
\end{align}
where $ \sigma^2_\scf = \textrm{Var}[E(R(t))]$ is the variance of the SCF error.
In practice, the variance $\sigma^2_\scf$ decays monotonically with
more strict SCF convergence criterion $\epsilon_\scf$.

The error estimates Eqs.~\eqref{eq:error-xpl} and \eqref{eq:error-scf} of the wavefunction prediction~\eqref{eq:error-decompose}
provide three qualitative guides on the choice of the extrapolation parameter $M$.
\begin{enumerate}
\item There always exists
an optimal choice of $M$ that minimizes the difference between the predicted and converged wavefunctions,
and thus minimizes the number of SCF iterations.
This is because as $M$ increases, the magnitude of $\mathcal E^\xpl$ decreases exponentially,
while the magnitude of $\mathcal E^\scf$ increases exponentially.
Thus the optimal $M$ is achieved at the cross-over of these two errors.
In general, it is difficult to predict, \emph{a priori}, the optimal value of $M$.
The knowledge, at least for an equilibrated system, may be numerically obtained from short testing simulations,
because the number of SCF iterations is also in equilibrium and does not significantly change with respect to time.
\item 
The optimal value of $M$ appears earlier when using smaller time steps.
The convergence of $\mathcal E^\xpl$ becomes fast when using smaller time steps,
and the magnitude of $\mathcal E^\scf$ does not depend on the time step size,
so the cross-over of the two errors occurs earlier.
That is to say a smaller $M$ is preferred in simulations using smaller time steps.
\item 
A larger $M$ is preferred in simulations that use more strict SCF convergence criteria (i.e.~smaller $\epsilon_\scf$).
It is noticed that the magnitude of $\mathcal E^\scf$ is proportional to the standard deviation of the SCF error,
and  $\mathcal E^\xpl$ is independent with the SCF convergence criterion.
Therefore, smaller $\epsilon_\scf$ postpones the cross-over of the two errors, and the optimal $M$ appears at a larger value.
\end{enumerate}
The correctness of these qualitative guides will be checked by the numerical examples presented in Sec.~\ref{sec:test}.

\recheck{
  The stability of BOMD in presence of the incomplete SCF iterations can be analyzed by
  calculating the roots of the characteristic equations of the extrapolation schemes~\cite{arias92prl}.  
  In this type of analysis, the SCF error was linearized and characterized by
  the largest eigenvalue of the SCF response kernel~\cite{steneteg2010}.
  It was concluded that by using higher orders of extrapolation schemes,
  the region of the stable SCF response kernel decreases,
  which indicates that a more strict SCF convergence criterion should be used to guarantee the stability.
  What is predicted from our error estimate is consistent
  with the finding obtained from the stability analysis.
}

\section{Wavefunction alignment}\label{sec:align}
In this section, we denote two sets of wavefunctions from two successive MD time steps by 
$\{\,\ket{\Phi_1^{(0)}}, \,\ldots, \,\ket{\Phi_{N_\nb}^{(0)}} \,\}$ and
$\{\,\ket{\Psi_1^{(0)}}, \,\ldots, \,\ket{\Psi_{N_\nb}^{(0)}} \,\}$, respectively,
with overlap matrix elements
\[ S_{ij}^{(0)} \equiv \langle\Psi_i^{(0)} | \Phi_j^{(0)} \rangle, \quad 1\leq i,j\le N_\nb, \]
and suppose that $S^{(0)}$ is non-singular.
The wavefunctions can not be combined directly in general, since the change from
$\ket{\Phi_j^{(0)}}$ to $\ket{\Psi_j^{(0)}}$ ($j\in\{1,2,\ldots,N_\nb\}$)
is frequently discontinuous.
As discussed in \cite{mead92,arias92prb,pancharatnam56,berry87},
it may be caused by various reasons including phase indeterminacies,
band crossing, energy level splitting, etc.
From a mathematical perspective, we understand the essence of these reasons as
the change in the choice of orthonormal basis in the wavefunction subspace,
which is defined as the linear subspace spanned by $N_\nb$ wavefunctions.
In order to deal with this discontinuity problem,
the orthonormal bases needs to be adjusted by unitary transformations on the original ones:
\begin{equation*}
    \left\{\begin{array}{rcl}
        \ket{\Phi_j} &\equiv& \sum_{\nu=1}^{N_\nb} \ket{\Phi_{\nu}^{(0)}}
        (U_{\Phi})_{\nu j},
        \vspace{0.1cm}\\
        \ket{\Psi_j} &\equiv& \sum_{\nu=1}^{N_\nb} \ket{\Psi_{\nu}^{(0)}}
        (U_{\Psi})_{\nu j},
    \end{array}\right.
\end{equation*}
where $U_{\Phi}$ and $U_{\Psi}$ are both unitary matrices.
It is easy to derive that the overlap matrix after the transformations becomes
$S = U_{\Psi}^{\dagger} \,S^{(0)} U_{\Phi}$.

The APJ scheme~\cite{arias92prb} minimizes the distance between the transformed bases,
and computes the unitary transformations by 
\begin{equation}
    \label{eq:aa-trans}
    \left\{\begin{array}{l}
    {U_{\Phi}^\ala}^{\dagger} \, \left( {S^{(0)}}^{\dagger} S^{(0)} \right)\,
    U_{\Phi}^\ala = D, \\
    U_{\Psi}^\ala = S^{(0)} U_{\Phi}^\ala \,D^{-1/2},
    \end{array}\right.
\end{equation}
where the superscript ``$\ala$'' indicates APJ scheme, and $D$ is a diagonal matrix.
Detailed derivation of the transformations is provided in Appendix \ref{app:align}.

Mead, in a review paper on Berry phase \cite{mead92},  defined
two sets of kets as in phase or parallel
if their overlap matrix was both Hermitian and positive-definite,
and any two sets of kets with a non-singular overlap matrix
can be made parallel by a unitary transformation on a certain one of them.
If letting $U_{\Psi}^\alm = I$, then 
\begin{equation}
    \label{eq:am-trans}
    U_{\Phi}^\alm = ({ S^{(0)} }^{\dagger} \, S^{(0)})^{-1/2}
    \, { S^{(0)} }^{\dagger}
\end{equation}
is uniquely determined \cite{mead92}, where the superscript ``$\alm$'' denotes Mead's scheme
. As discussed in Appendix \ref{app:align},
$({ S^{(0)} }^{\dagger} \, S^{(0)})^{-1/2}$ can be implemented through
a singular value decomposition of ${S^{(0)}}^{\dagger}$
or a matrix diagonalization of ${ S^{(0)} }^{\dagger} \, S^{(0)}$,
and the latter is computationally more effective.

In the APJ scheme, the overlap matrix after transformation
${U_{\Psi}^\ala}^{\dagger} \,S^{(0)} U_{\Phi}^\ala$ is a diagonal matrix
with positive elements (see Appendix \ref{app:align}).
Consider that we keep 
$\{\,\ket{\Psi_1^{(0)}}, \,\ldots, \,\ket{\Psi_{N_\nb}^{(0)}} \,\}$
unchanged, and transform $\{\,\ket{\Phi_1^{(0)}}, \,\ldots, \,\ket{\Phi_{N_\nb}^{(0)}} \,\}$
using $U_{\Phi}^\ala \,{U_{\Psi}^\ala}^{\dagger}$.
Then the overlap matrix after transformation
\[ S = S^{(0)} U_{\Phi}^\ala \,{U_{\Psi}^\ala}^{\dagger} = S^{(0)} U_{\Phi}^\ala
D^{-1/2} {U_{\Phi}^\ala}^{\dagger} {S^{(0)}}^{\dagger} \]
is obviously both Hermitian and positive-definite.
According to the uniqueness of $U_{\Phi}^\alm$, we conclude that
\begin{equation}
    \label{eq:aa-am}
    U_{\Phi}^\ala \,{U_{\Psi}^\ala}^{\dagger} =U_{\Phi}^\alm.
\end{equation}
By using~\eqref{eq:aa-am},
it is straightforward to show that the extrapolated wavefunctions
using these two alignment schemes 
differ only by a unitary transformation:
\begin{equation}
  \label{eq:xtrpl-a-m}
\left( X_1^{\mathrm{p}},\,\ldots,\,X_{N_\nb}^{\mathrm{p}} \right)^\ala =
\left( X_1^{\mathrm{p}},\,\ldots,\,X_{N_\nb}^{\mathrm{p}} \right)^\alm \, U_{\Psi}^\ala.  
\end{equation}
Noticing that the extrapolated wavefunctions should be orthonormalized before using as the initial guess for the SCF iteration,
the theoretical comparison of extrapolation accuracies by using the two alignment schemes
is indeed a subtle problem.
The numerical performance of these two alignment schemes are similar due to the presented equivalency~\eqref{eq:xtrpl-a-m}, and will be discussed in Sec.~\ref{sec:test}.

\section{Numerical results}\label{sec:test}
We have implemented the TX and GX schemes mentioned in Sec.~\ref{sec:txgx}
and also the APJ and Mead's wavefunction alignment schemes in Sec.~\ref{sec:align},
in an in-house \textit{ab-initio} simulation package CESSP
developed on infrastructure JASMIN \cite{mo10},
and carried out systematic tests on a solid copper system with 64 atoms to validate our theoretical analysis.
At each MD time step, the Kohn--Sham equation under
the generalized gradient approximation (GGA) \cite{perdew96}
was solved by the plane-wave pseudopotential method \cite{payne92,kresse96cms,martin04}.
The kinetic energy cutoff for the plane-wave expansion was chosen as 350\,eV,
and a $2\times 2\times 2$ $k$-points mesh was used.
In the SCF iteration, the Pulay's mixing scheme \cite{pulay80} was employed to
mix the electronic charge densities,
and the SCF convergence was considered to be reached when differences in
both the potential energy and band structure energy between successive SCF steps
are smaller than the criterion $\epsilon_\scf$.
The iterative matrix diagonalization of the Kohn--Sham Hamiltonian was
realized by the block Davidson scheme \cite{liu78}.

The initial coordinates and velocities were obtained by a warm-up microcanonical simulation,
which started from a perfect FCC configuration with atomic velocities sampled from a Maxwell-Boltzmann distribution at temperature 1400\,K.
The system was equilibrated after 1\,ps to around 700\,K and the FCC lattice was preserved.
Then we launched all productive simulations from the same initial coordinates and velocities.
The integration of the equation of motion was implemented using the
Verlet scheme~\cite{verlet1967computer}
and the tested time step sizes were chosen to be $\Delta t = 0.5$, 1.0, 2.0, and 4.0\,fs.
In the productive simulations, since the system was equilibrated,
the number of SCF iterations did not significantly change with respect to time,
we thus computed the time-average of the number of SCF iterations
and used it to calibrate the effectiveness of wavefunction extrapolation.

\begin{table}   
    \centering
    \caption{The averaged numbers of SCF iterations and total energy drift results
    of the TX scheme under different time step sizes.
    The SCF convergence criterion is $\epsilon_\scf = 10^{-5}$\,eV.
    \#SCF: Averaged number of SCF iterations, excluding the first $M$ time steps.
    Drift (in eV/ps/atom): Absolute value of the linearly fitted total energy drift.}
    \label{tbl:cu64_tp_ts}
    \begin{tabular}{l|ll|ll|ll|ll}
        \hline
        & \multicolumn{2}{c|}{$\Delta t$ = 0.5\,fs}
        & \multicolumn{2}{c|}{$\Delta t$ = 1.0\,fs}
        & \multicolumn{2}{c|}{$\Delta t$ = 2.0\,fs}
        & \multicolumn{2}{c}{$ \Delta t$ = 4.0\,fs} \\\hline
        $M$ & \#SCF & Drift & \#SCF & Drift & \#SCF & Drift & \#SCF & Drift \\\hline
        2 & 3.36 & 1.80e-03 & 5.12 & 1.66e-03 & 7.00 & 9.64e-04 & 9.69 & 1.51e-03 \\
        3 & 4.11 & 1.00e-03 & 4.27 & 7.98e-03 & 4.45 & 1.21e-03 & 7.98 & 3.75e-03 \\
        4 & 4.76 & 1.43e-03 & 5.16 & 1.14e-03 & 5.67 & 2.22e-04 & 6.87 & 5.52e-04 \\
        5 & 6.01 & 1.17e-03 & 6.22 & 1.52e-03 & 6.61 & 9.35e-04 & 6.96 & 9.91e-04 \\
        \hline
    \end{tabular}
\end{table}

\begin{table}
    \centering
    \caption{The averaged numbers of SCF iterations and total energy drift results
    of the TX scheme under different SCF convergence criteria ($\Delta t$ = 2.0\,fs).
    \#SCF and Drift (in eV/ps/atom) have the same meanings as in Tab.~\ref{tbl:cu64_tp_ts}.}
    \label{tbl:cu64_tp_scf}
    \begin{tabular}{l|ll|ll|ll|ll|ll|ll}
        \hline
        & \multicolumn{2}{c|}{$\epsilon_\scf$ = $10^{-3}$\,eV}
        & \multicolumn{2}{c|}{$\epsilon_\scf$ = $10^{-4}$\,eV}
        & \multicolumn{2}{c|}{$\epsilon_\scf$ = $10^{-5}$\,eV}
        & \multicolumn{2}{c|}{$\epsilon_\scf$ = $10^{-6}$\,eV}
        & \multicolumn{2}{c|}{$\epsilon_\scf$ = $10^{-7}$\,eV}
        & \multicolumn{2}{c}{$ \epsilon_\scf$ = $10^{-8}$\,eV} \\\hline
        $M$ & \#SCF & Drift & \#SCF & Drift & \#SCF & Drift
        & \#SCF & Drift & \#SCF & Drift & \#SCF & Drift \\\hline
        2 & 3.08 & 2.01e-02 & 5.00 & 4.61e-03 & 7.00 & 9.64e-04 & 9.20 & 3.01e-04 & 10.87 & 1.44e-04 & 14.15 & 2.72e-06 \\
        3 & 3.48 & 5.50e-02 & 4.06 & 2.79e-02 & 4.45 & 1.21e-03 & 6.26 & 2.14e-04 &  8.99 & 5.94e-05 & 12.17 & 1.46e-05 \\
        4 & 4.35 & 1.33e-02 & 4.99 & 3.22e-03 & 5.67 & 2.22e-04 & 5.80 & 1.11e-04 &  6.25 & 7.25e-05 &  9.55 & 8.01e-06 \\
        5 & 5.78 & 6.46e-03 & 6.06 & 4.67e-03 & 6.61 & 9.35e-04 & 6.99 & 2.07e-04 &  7.07 & 1.08e-04 &  7.42 & 5.68e-05 \\
        \hline
    \end{tabular}
\end{table}

To validate the theoretical analysis presented in Sec.~\ref{sec:err},
we tested TX scheme with $M$ varying from 2 to 5.
Firstly, we fixed the SCF convergence criterion to $\epsilon_\scf$ = $10^{-5}$\,eV,
and carried out simulations using different time step sizes.
The averaged numbers of SCF iterations are shown in Tab.~\ref{tbl:cu64_tp_ts}.
It is observed from the table that for any time step choice,
the number of SCF iterations firstly decreases,
and then increases by using an increasing value of $M$
(for time step $\Delta t$ = 0.5\,fs and $M=1$,
the average number of SCF iterations is 6.72 that is not shown in the table).
The optimal value of $M$ appears at $M=2$, 3, 3, and 4 for time step $\Delta t = 0.5$, 1.0, 2.0 and 4.0~fs, respectively.
From this observation, it is concluded that the optimal $M$ in terms of minimal number of SCF iterations
appears at larger values when using larger time steps,
and this trend is consistent with the theoretical analysis on the optimal $M$ value presented in Sec.~\ref{sec:err}.
Results in Tab.~\ref{tbl:cu64_tp_scf} illustrate that when fixing the time step,
a more strict $\epsilon_\scf$ will lead to a larger optimal value of $M$:
For convergence criterion $\epsilon_\scf = 10^{-3}$\,eV, the optimal $M$ is as small as 2
(when $M=1$ the number of SCF iterations is 5.06 that is not shown in the table),
while it takes as large as 5 for convergence criterion $\epsilon_\scf = 10^{-8}$
(when $M=6$ the number of SCF iterations is 8.35 that is not shown in the table).
This trend is also consistent with the theoretical analysis in Sec.~\ref{sec:err}.

Observed from the numerical results in Tabs.~\ref{tbl:cu64_tp_ts} and \ref{tbl:cu64_tp_scf},
the value $M=3$ is only optimal for $\Delta t = 1.0\,\textrm{fs},\ \epsilon_\scf = 10^{-5}$~eV;
$\Delta t = 2.0\,\textrm{fs},\ \epsilon_\scf = 10^{-4}$~eV and
$\Delta t = 2.0\,\textrm{fs},\ \epsilon_\scf = 10^{-5}$~eV.
The other values of $M$ (either smaller or larger) should be used in other cases
to minimize the number of SCF iterations.
Taking the case of $\Delta t = 2.0\,\textrm{fs},\ \epsilon_\scf = 10^{-8}$~eV for example,
the optimal value $M=5$ saves about 39\% SCF iterations comparing with $M=3$.
Therefore, $M=3$, as the default setting provided in most of the popular AIMD packages
(see Tab.~\ref{tbl:pkginfo}), may be significantly slower than the optimal $M$. 

As a reference, we also provide the energy drift results in Tabs.~\ref{tbl:cu64_tp_ts} and \ref{tbl:cu64_tp_scf}.
They were evaluated by fitting the total energies to a linear function of time.
This is a reasonable ansatz because the time evolution of the total energy drift presents
a linear dependence on the simulation time,
which is consistent with earlier results on the energy conservation of BOMD
reported in e.g.~Refs.~\cite{marx-hutter00,pf04,herbert05}.
It should be noted that neither the TX nor GX scheme is designed
to conserve the total energy of the system, therefore, not surprisingly,
a systematic energy drift is observed in all the numerical tests.
Moreover, the numerical results show that
increasing $M$ or decreasing $\Delta t$ does not guarantee a smaller drift, 
while tightening the SCF convergence criterion $\epsilon_\scf$ can effectively improve the energy conservation.
\recheck{
  The energy drift,
  which cannot be entirely removed even when using a very strict SCF criterion (e.g.~$\epsilon_\scf = 10^{-8}$~eV),  
  is considered to stem from a broken time reversal symmetry
  along with the implicit electronic propagation composed of
  the extrapolation step and the SCF procedure~\cite{kuhne07,niklasson06}.
  New methods have thus been developed to restore the time reversibility,
  for example,
  the ASPC (Always Stable Predictor--Corrector)~\cite{kolafa04,kuhne07} scheme 
  and the extended Lagrangian BOMD
  (XL-BOMD)~\cite{niklasson06,niklasson08,niklasson09,niklasson12,steneteg2010} method.
  In particular, the XL-BOMD method is shown to precisely conserve the total energy.
}


\begin{table}[t]
  \centering
  \caption{The averaged extrapolation coefficients of the $M=4$ GX scheme.
    The SCF convergence criterion is $\epsilon_\scf = 10^{-5}$~eV.
    The TX coefficients (constants) are presented as a comparison.
  }
  \label{tbl:cu64_gx_ck}
  {\setlength{\tabcolsep}{1em}
  \begin{tabular}{c|c|c|c|c}
    \hline 
    Scheme              &$\Delta t$ (fs)    & $c_1$ & $c_2$ & $c_3$ \\
    \hline
    \multirow{4}{*}{GX} &0.5                & 2.99  & $-2.99$ & 1.00  \\
                        &1.0                & 2.99  & $-2.99$ & 1.00  \\
                        &2.0                & 2.99  & $-2.98$ & 0.99  \\
                        &4.0                & 2.95  & $-2.92$ & 0.97 \\ \hline
    TX                  & --                & 3.00  & $-3.00$ & 1.00 \\
    \hline 
  \end{tabular}}
\end{table}

\begin{table}[t]
    \centering
    \caption{The averaged numbers of SCF iterations under the GX scheme,
    excluding the first $M$ time steps.
    The SCF convergence criterion is $\epsilon_\scf = 10^{-5}$\,eV.}
    \label{tbl:cu64_g}
    \begin{tabular}{l|l|l|l|l}
        \hline
        $M$ & $\Delta t$ = 0.5\,fs & $\Delta t$ = 1.0\,fs
            & $\Delta t$ = 2.0\,fs & $\Delta t$ = 4.0\,fs \\\hline
        2 & 3.35 & 5.11 & 7.02 & 9.69 \\
        3 & 4.09 & 4.03 & 4.34 & 7.96 \\
        4 & 4.73 & 5.17 & 5.65 & 6.27 \\
        5 & 6.01 & 6.21 & 6.35 & 6.86 \\
        \hline
    \end{tabular}
\end{table}

It has been remarked in Sec.~\ref{sec:txgx} that extrapolation coefficients of the GX scheme
are in fact close to those of the TX scheme, especially for small time step sizes.
To test this point, we have carried out a set of similar tests using the GX scheme for
time step $\Delta t = 0.5$, 1.0, 2.0 and 4.0\,fs and SCF convergence criterion $\epsilon_\scf = 10^{-5}$~eV.
The extrapolation coefficients of the $M=4$ case are listed in Tab.~\ref{tbl:cu64_gx_ck}.
It is observed that the TX and averaged GX coefficients are very similar.
Even for the case of $\Delta t = 4.0$\,fs,
when the difference between the two schemes is expected to be large,
the maximal deviation in coefficients is only 3\%.
The averaged numbers of SCF iterations of $M=2$ to 5 are provided in Tab.~\ref{tbl:cu64_g}.
A comparison with Tab.~\ref{tbl:cu64_tp_ts} that lists the averaged numbers of SCF iterations
of the TX scheme illustrates a clear similarity between the GX and TX schemes.

\begin{table}[t]
    \centering
    \caption{The averaged numbers of SCF iterations under
    APJ wavefunction alignment and TX schemes, excluding the first $M$ time steps.
    The SCF convergence criterion is $\epsilon_\scf = 10^{-5}$\,eV.}
    \label{tbl:cu64_tp_arias}
    \begin{tabular}{l|l|l|l|l}
        \hline
        $M$ & $\Delta t$ = 0.5\,fs & $\Delta t$ = 1.0\,fs
            & $\Delta t$ = 2.0\,fs & $\Delta t$ = 4.0\,fs \\\hline
        2 & 3.16 & 4.69 & 8.77 & 10.28  \\
        3 & 3.21 & 3.33 & 4.63 &  8.93  \\
        4 & 5.52 & 5.53 & 5.37 &  6.19  \\
        5 & 7.12 & 7.26 & 7.17 &  7.05  \\
        \hline
    \end{tabular}
\end{table}

By far all tests have employed the Mead's wavefunction alignment scheme,
as a comparison, the averaged numbers of SCF iterations using the APJ alignment scheme for
time step $\Delta t = 0.5$, 1.0, 2.0 and 4.0~fs and SCF convergence criterion $\epsilon_\scf = 10^{-5}$~eV
are presented in Tab.~\ref{tbl:cu64_tp_arias}.
A comparison between Tabs. \ref{tbl:cu64_tp_ts} and \ref{tbl:cu64_tp_arias} illustrates that 
switching to APJ alignment scheme does not change the optimal value of $M$,
which remains as $M=2$, 3, 3, and 4 for time step $\Delta t = 0.5$, 1.0, 2.0 and 4.0~fs, respectively.
Moreover, considering the optimal number of SCF iterations,
it is observed that none of the two alignment schemes presents a systematic advantage over
the other, and the differences between them are in fact always less than one.
We understand these results as a reflection of the theoretical equivalence
(ignoring a unitary transformation) between the extrapolated wavefunctions by using the two alignment schemes.



\begin{figure}
    \centering
    \includegraphics[width=0.48\textwidth]{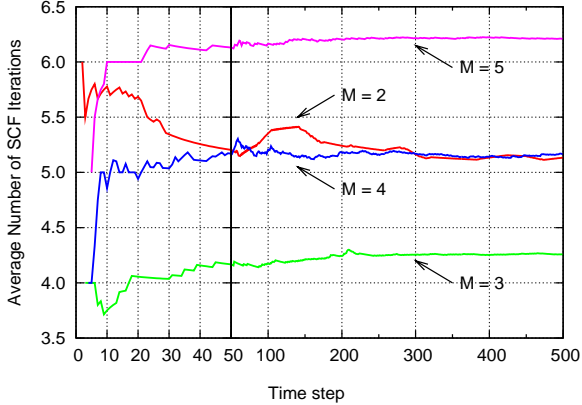}
    \caption{The averaged numbers of SCF iterations calculated along the MD trajectories
    under $M=2$ to 5 with time step $\Delta t$ = 1.0\,fs.
    The value for time step $n$ is averaged from the $M$-th to the $n$-th time steps.
    The SCF convergence criterion is $\epsilon_\scf = 10^{-5}$\,eV,
    and the Mead's alignment and TX scheme are used.
    }
    \label{fig:scfavg}
\end{figure}

In the above we have demonstrated that the value of $M$ is most relevant to the number of SCF iterations, and 
the choice between the two wavefunction alignment schemes
or between the TX and GX schemes does not result in a significant difference in the
effectiveness of wavefunction extrapolation.
In practical simulations of equilibrium systems, the optimal $M$ may be obtained by short testing simulations, 
because we notice that  in all the testing simulations 
the average number of SCF iterations calculated along the MD trajectory
enters a steady state only after a few tens of time steps.
As an example, the case of $\Delta t = 1.0$~fs and $\epsilon_\scf = 10^{-5}$~eV is shown in Fig.~\ref{fig:scfavg}.

\section{Conclusions and discussions}\label{sec:sum}
In this work,
we investigated both theoretically and numerically the optimal choice of
the critical parameter $M$ in the wavefunction extrapolation, which denotes the number of previous MD time steps used to construct the initial guess for the SCF iteration.
We focused on two types of extrapolations schemes, i.e.~time extrapolation (TX) and geometric extrapolation (GX)
that are most frequently implemented in the popular AIMD simulation packages using the Kohn--Sham density functional theory
and a plane-wave basis set.

We showed that the extrapolation error is divided into two parts:
The first stems from extrapolating the fully converged wavefunctions and decays exponentially with respect to the parameter $M$.
The second is a linear combination of the SCF errors and explodes exponentially with respect to the parameter $M$.
Therefore, an optimal value of $M$ always exists.
Since the convergence speed of the first term is governed by the time step and the magnitude of the second term is proportional to the SCF error,
we qualitatively predicted that the optimal $M$ tends to get larger when
increasing the MD time step size or tightening the SCF convergence criterion.
Moreover, we theoretically explained the reason why the TX and GX schemes have much similarity,
therefore the analysis for the TX scheme can be extended to the GX scheme without substantial difficulty.
A solid copper system with 64 atoms was chosen to test our theoretical statements.
Numerical results on the appearance of optimal values of $M$ for the TX scheme
under a set of different time step sizes and SCF convergence criteria well confirmed the trends predicted by the error analysis.
Also, a comparison between TX and GX validated their similarity.
It was demonstrated that the optimal value of $M$ takes 2 in some cases, and
takes as large as 5 in other cases, which saves about 39\% SCF iterations comparing with $M=3$
(the default setting in most popular AIMD packages). 
Therefore, our work suggest the necessity for AIMD simulation packages to
open the user interface and provide more choices on the parameter $M$.
\recheck{
  It should be noted that
  the extrapolation schemes need the storage of the wavefunctions from the previous $M$ time steps,
  so the additional memory cost grows linearly with respect to an increasing $M$.
  Therefore, in practice, the user should also consider the balance between the computational and
  memory costs when choosing the value of $M$.
}

For the prerequisite alignment step, we theoretically proved that the extrapolated wavefunctions
by using APJ and Mead's schemes differ only by a unitary transformation,
and illustrated numerically the essential equivalence between them.
The only concern is for the numerical implementation,
that a matrix diagonalization of ${ S^{(0)} }^{\dagger} \, S^{(0)}$ other than
a singular value decomposition of ${S^{(0)}}^{\dagger}$
is recommended due to the computational efficiency.

Although the TX and GX schemes could significantly accelerate SCF convergence,
it should be noted that neither of them is designed for total energy conservation.
Correspondingly, a systematic total energy drift in microcanonical trajectories exists,
and increasing $M$ or decreasing the time step does not guarantee a smaller drift. 
Therefore, in the case when the energy conservation is critical,
we suggest either a very strict SCF convergence criterion,
or new techniques like the ASPC or XL-BOMD method.

In the end, we would like to point out that
our work only prove the existence of the optimal value $M$, and do not tell, \emph{a priori},
\recheck{
  how large $M$ is for a specific system and how much the advantage is by using the optimal $M$ instead of the widely accepted value of~3.
  We notice that 
  for the tested copper system,   $M=3$ is already optimal
  under the normally used simulation setups: A time step of $1\sim 2$~fs
  and a SCF convergence criterion of $10^{-4}\sim 10^{-5}$~eV.
  However, this experience 
  cannot be directly extended to other systems.
  Therefore the value of this work is to point out the existence of the optimal $M$,
  and the possibility of improving the extrapolation quality,
  marginally or dramatically, by tuning the value of $M$.
  In the simulations of equilibrium systems, the optimal $M$ can be estimated by short testing simulations of a few dozen time steps.
  Moreover, the direction of changing $M$ is also predicted by our theory when using a different time step size or SCF convergence criterion.}
In the simulations of non-equilibrium systems, the optimal choice of $M$ is still an open question, and will be investigated in the future.

\section{Acknowledgment}
The authors acknowledge the valuable discussions with Dr.~Deye Lin,
and gratefully acknowledge the financial support from
National High Technology Research and Development Program of China under Grant 2015AA01A304.
X.G. is supported by the National Science Foundation of China under Grants 91430218 and 61300012.
H.W. is supported by the National Science Foundation of China under Grants 11501039 and 91530322.

\appendix

\section{Details about wavefunction alignment}\label{app:align}
\subsection{Derivation of $U_{\Phi}^\ala$ and $U_{\Psi}^\ala$}
We need to solve the following constrained minimization problem:
\begin{equation}
    \label{eq:aa-problem}
    \min_{U_{\Phi}^\ala,\,U_{\Psi}^\ala} f \equiv
    \sum_{j=1}^{N_\nb} w_j \Big\| \ket{\Psi_j} -\ket{\Phi_j} \Big\|^2
    = \sum_{j=1}^{N_\nb} w_j \int_{\Omega}
    \Big|\Psi_j(\vec{r}) - \Phi_j(\vec{r})\Big|^2 d\vec{r},
\end{equation}
where
$\ket{\Phi_j} = \sum_{\nu=1}^{N_\nb} \ket{\Phi_{\nu}^{(0)}}(U_{\Phi}^\ala)_{\nu j}$
and
$\ket{\Psi_j} = \sum_{\nu=1}^{N_\nb} \ket{\Psi_{\nu}^{(0)}}(U_{\Psi}^\ala)_{\nu j}$
are unitary transformed functions from two sets of wavefunctions
given by successive MD time steps, $w_j>0$ the weight of the $j$-th function,
and $\Omega$ the computational domain.
It is convenient to rewrite objective function $f$ in matrix form:
\[ f = \mathrm{tr} [W (2I - S - S^{\dagger}) ]
= \mathrm{tr} \left[W \left(2I -{U_{\Psi}^\ala}^{\dagger}\,S^{(0)}\,U_{\Phi}^\ala -
{U_{\Phi}^\ala}^{\dagger}\,{ S^{(0)} }^{\dagger} U_{\Psi}^\ala\right) \right], \]
where tr represents the trace of a matrix, $W\equiv\mathrm{diag}\{w_1,\ldots,w_{N_\nb}\}$,
$I$ is the identity matrix, and $S^{(0)}$ and $S$ are overlap matrices before and after
the unitary transformations, respectively.

We introduce additional terms
$\mathrm{tr} [\Lambda_{\Phi} ({U_{\Phi}^\ala}^{\dagger}U_{\Phi}^\ala - I)]$
and
$\mathrm{tr} [\Lambda_{\Psi} ({U_{\Psi}^\ala}^{\dagger}U_{\Psi}^\ala - I)]$
to impose the constraints that $U_{\Phi}^\ala$ and $U_{\Psi}^\ala$
are unitary matrices, and obtain Lagrange function
\[ L =
\mathrm{tr}[W(2I -{U_{\Psi}^\ala}^{\dagger}\,S^{(0)} U_{\Phi}^\ala
- {U_{\Phi}^\ala}^{\dagger}\,{S^{(0)}}^{\dagger} U_{\Psi}^\ala)]
+ \Lambda_{\Phi}({U_{\Phi}^\ala}^{\dagger}U_{\Phi}^\ala - I)
+ \Lambda_{\Psi}({U_{\Psi}^\ala}^{\dagger}U_{\Psi}^\ala - I)],
\]
where Lagrange multiplier matrices $\Lambda_{\Phi}$ and $\Lambda_{\Psi}$ are
both Hermitian. Taking partial derivatives of $L$ with respect to
real and imaginary parts of matrix elements of $U_{\Phi}^\ala$ and
$U_{\Psi}^\ala$, we get
\begin{equation}
    \label{eq:aa-eq}
    \left\{\begin{array}{rcl}
        {U_{\Phi}^\ala}^{\dagger}\, {S^{(0)}}^{\dagger} U_{\Psi}^\ala &=&
        \Lambda_{\Phi} W^{-1}, \\
        {U_{\Psi}^\ala}^{\dagger}\, S^{(0)} U_{\Phi}^\ala &=& \Lambda_{\Psi} W^{-1}.
    \end{array}\right.
\end{equation}

It is easy to see from (\ref{eq:aa-eq}) that
\[ \left\{\begin{array}{rcl}
    \Lambda_{\Phi} W &=& W \Lambda_{\Psi}, \\
    W \Lambda_{\Phi} &=& \Lambda_{\Psi} W,
\end{array}\right. \]
and then
\[ \left\{\begin{array}{rcl}
    \Lambda_{\Phi} W^2 &=& W \Lambda_{\Psi} W = W^2 \Lambda_{\Phi}, \\
    \Lambda_{\Psi} W^2 &=& W \Lambda_{\Phi} W = W^2 \Lambda_{\Psi},
\end{array}\right. \]
which means $\Lambda_{\Phi}$ and $\Lambda_{\Psi}$ are both commutative with $W^2$.
Since this is true for any $W=\mathrm{diag}\{w_1,\ldots,w_{N_\nb}\}~(w_j > 0)$,
we assume that any two matrix elements of $W$ are not equal,
and conclude that $\Lambda_{\Phi}$ and $\Lambda_{\Psi}$ are diagonal matrices.
Thus, the second equation in (\ref{eq:aa-eq}) implies that
\begin{equation*}
    {U_{\Phi}^\ala}^{\dagger} \, \left( {S^{(0)}}^{\dagger} S^{(0)} \right)\,
    U_{\Phi}^\ala
    = (\Lambda_{\Psi}W^{-1})^2 \equiv D,
\end{equation*}
i.e., $U_{\Phi}^\ala$ is determined as the unitary matrix
that diagonalizes ${S^{(0)}}^{\dagger}\,S^{(0)}$.
Further, according to (\ref{eq:aa-eq}) we have
\[ U_{\Psi}^\ala = S^{(0)} U_{\Phi}^\ala (\Lambda_{\Psi}W^{-1})^{-1}
= \pm \,S^{(0)} U_{\Phi}^\ala D^{-1/2}. \]
The remained work is to determine the sign. Recalling
$S = {U_{\Psi}^\ala}^{\dagger} S^{(0)} U_{\Phi}^\ala$,
the second equation in (\ref{eq:aa-eq}) is in fact $S = \Lambda_{\Psi} W^{-1}$.
Then the objective function
\[ f = \mathrm{tr} [W (2I - S - S^{\dagger}) ] = 2 \sum_{j=1}^{N_\nb} w_j (1-s_j). \]
Because $|s_j| = |\langle\Psi_j|\Phi_j\rangle| \le 1$,
$f$ reaches its minimum when $s_j\ge 0$. Thus, we get
\begin{equation*}
    U_{\Psi}^\ala = S^{(0)} U_{\Phi}^\ala \,D^{-1/2}.
\end{equation*}

\subsection{The way to compute $U_{\Phi}^\alm$}
One choice is to carry out singular value decomposition of ${S^{(0)}}^{\dagger}$:
\begin{equation}
    \label{eq:svd}
    {S^{(0)}}^{\dagger} = U_1 \,\Sigma \,U_2^{\dagger},
\end{equation}
where $U_1$ and $U_2$ are both unitary matrices,
and $\Sigma$ is a diagonal matrix with non-negative elements.
Since $S^{(0)}$ is assumed to be non-singular, matrix elements of $\Sigma$ is in fact positive.
Then
\[ U_{\Phi}^\alm
= ({ S^{(0)} }^{\dagger} \, S^{(0)})^{-1/2} \,{ S^{(0)} }^{\dagger}
= (U_1 \,\Sigma^2 \, U_1^{\dagger})^{-1/2} \, U_1 \,\Sigma \,U_2^{\dagger}.\]
It is easy to see that $ (U_1 \,\Sigma^2 \, U_1^{\dagger})^{-1/2}
= U_1 \,\Sigma^{-1} \, U_1^{\dagger}$, so we have
\[ U_{\Phi}^\alm = U_1 \, U_2^{\dagger}. \]

The other choice is matrix diagonalization:
\[ U^{\dagger} \left( {S^{(0)}}^{\dagger} S^{(0)} \right) U = D, \]
where $U$ is a unitary matrix and $D$ is a diagonal matrix with positive elements.
It is easy to see that
\[ \left( {S^{(0)}}^{\dagger} S^{(0)} \right)^{-1} = U D^{-1} \, U^{\dagger}. \]
The R.H.S. of this equation can be written as
$U D^{-1/2} \, U^{\dagger} \,U D^{-1/2} \, U^{\dagger}$, so
\[ \left( {S^{(0)}}^{\dagger} S^{(0)} \right)^{-1/2} = U D^{-1/2} \, U^{\dagger}, \]
and
\[ U_{\Phi}^\alm = U D^{-1/2} \,U^{\dagger} \,{S^{(0)}}^{\dagger}.\]

Since singular value decomposition is more computationally consuming than
diagonalization, we suggest the latter choice in numerical implementation.


\section{Estimate the error $\mathcal E^\scf$}
\label{app:prove-error-scf}
In this section we prove Eq.~\eqref{eq:error-scf}.
We assume that the SCF errors $E(R(t_n))$ are independent random variables that have vanishing mean and identical variance denoted by $\sigma^2_\scf$.
As a short-hand notation, we write $E_n = E(R(t_n))$.
Noticing Eq.~\eqref{eq:decomp-error-scf}, it is clear that the mean of $\mathcal E^\scf$ vanishes,
and the variance of $\mathcal E^\scf$ is estimated by
\begin{align*}
  \langle (\mathcal E^\scf)^2\rangle
  = &\,
  \Big\langle
  \Big[ E_n + \sum_{k=1}^{M-1} c_k E_{n-k+1} - \sum_{k=1}^{M-1} c_k  E_{n-k} \Big]^2
  \Big\rangle   \\
  = &\,
  \Big\langle
  E_n^2
  + \Big(\sum_{k=1}^{M-1} c_k E_{n-k+1}\Big)^2
  + \Big(\sum_{k=1}^{M-1} c_k E_{n-k}\Big)^2 \\ 
  &
  + 2 c_1E_n^2
  - 2 \Big(\sum_{k=1}^{M-1} c_k E_{n-k+1}\Big) \Big(\sum_{k=1}^{M-1} c_k E_{n-k}\Big)
  \Big\rangle \\ 
  = &\,
  \Big\langle
  E_n^2
  + \sum_{k=1}^{M-1} c_k^2 E^2_{n-k+1}
  + \sum_{k=1}^{M-1} c_k^2 E^2_{n-k}
  + 2 c_1E_n^2
  - 2 \sum_{k=1}^{M-2}c_kc_{k+1}E^2_{n-k}
  \Big\rangle\\ 
  = &\,
  \sigma_\scf^2
  \Big(
  1 + 2 c_1 + 2 \sum_{k=1}^{M-1} c_k^2  - 2  \sum_{k=1}^{M-2}c_kc_{k+1}
  \Big),
\end{align*}
where we used the fact that $\langle E_n\rangle = 0$ and $\langle E_m E_n\rangle = \delta_{mn}\,\sigma^2_\scf$.
Noticing that $c_k = (-1)^{k-1} \binom{M-1}{k}$, we reach
\begin{align*}
  \langle (\mathcal E^\scf)^2\rangle
  = &\,
  \sigma_\scf^2
  \Bigg[
  1 + 2 (M-1)
  + 2 \sum_{k=1}^{M-1} \binom{M-1}{k}^2
  + 2 \sum_{k=1}^{M-1}\frac{M-1-k}{k+1} \binom{M-1}{k}^2
  \Bigg]   \\
  = &\,
  \sigma_\scf^2
  \Bigg[
  - 1
  + 2 \sum_{k=0}^{M-1}\frac{M}{k+1} \binom{M-1}{k}^2
  \Bigg].
\end{align*}
The magnitude of $\langle (\mathcal E^\scf)^2\rangle$ can be estimated by
\begin{align*}
  \langle (\mathcal E^\scf)^2\rangle
  & >
  \sigma_\scf^2
  \Bigg[
  - 1
  + 2 \sum_{k=0}^{M-1}\binom{M-1}{k}^2
  \Bigg]  \\
  &\geq
  \sigma_\scf^2
  \Bigg\{
  - 1
  + \frac{2}{M}
  \Bigg[
  \sum_{k=0}^{M-1} \binom{M-1}{k}
  \Bigg]^2
  \Bigg\} \\
  &=
  \sigma_\scf^2
  \Big\{
  - 1
  + \frac{2^{(2M-1)}}{M} 
  \Big\}.  
\end{align*}
Therefore, the growing speed of the standard deviation of the extrapolated SCF noise is lower bounded by
\begin{align}
  \sqrt{\langle (\mathcal E^\scf)^2\rangle} > \frac{2^{M-1}}{\sqrt {M}} \sigma_\scf.
\end{align}


\end{document}